\documentclass[prl,reprint]{revtex4-1}
\usepackage{amsmath}
\usepackage{amssymb}
\usepackage{graphicx}
\usepackage{textcomp}
\begin{document}

\title{Practical x-ray ghost imaging with synchrotron light}

%

\author{Daniele Pelliccia}
\email[]{daniele@idtools.com.au}
\affiliation{Instruments \& Data Tools Pty Ltd, Victoria 3178, Australia}
\affiliation{School of Science, RMIT University, Victoria 3001, Australia}
\author{Margie P. Olbinado}
\affiliation{European Synchrotron Radiation Facility, 38043 Grenoble, France}
\author{Alexander Rack}
\affiliation{European Synchrotron Radiation Facility, 38043 Grenoble, France}
\author{David M. Paganin}
\affiliation{School of Physics and Astronomy, Monash University, Victoria 3800, Australia}



\begin{abstract}
We present a practical experimental realization of transmission x-ray ghost imaging using synchrotron light. Hard x-rays from an undulator were split by a Si 200 crystal in Laue geometry to produce two copies of a speckled incident beam. Both speckle beams were simultaneously measured on a CCD camera. The sample was inserted in one of the two beams, and the corresponding image was integrated over the camera extent to synthesize a bucket signal. We show the successful x-ray ghost image reconstruction of two samples and discuss different reconstruction strategies. We also demonstrate a method for measuring the point spread function of a ghost imaging system, which can be used to quantify the resolution of the ghost imaging reconstructions and quantitatively compare different reconstruction approaches. Our experimental results are discussed in view of future practical applications of x-ray ghost imaging, including a means for parallel ghost imaging. 
\end{abstract}

\maketitle 
\section{Introduction}
Ghost imaging \cite{aa, aaa, a} was first proposed \cite{GI-proposal-with-visible-light-1, GI-proposal-with-visible-light-2} and then experimentally achieved \cite{GI-expt-with-visible-light, GI-expt-with-visible-light-3} in visible-light quantum optics.  It utilizes intensity correlations between (i) spatially resolved photons which never pass through a sample of interest and (ii) non-spatially resolved photons that do pass through the sample. It is remarkable that this parallelized form of the Hanbury Brown--Twiss experiment \cite{b1, b2, Wang2009} permits images of a sample to be obtained, despite the fact that photons passing through the object are detected using only a single ``bucket'' detector \cite{c, d}.  Interestingly, the use of delocalized photons created by passage through a beam-splitter, in the standard quantum-optics setup for ghost imaging, negates the classical dualism of a spatially confined photon either locally interacting with an object or locally not interacting with an object.  Accordingly, there has been much debate as to whether ghost imaging is or is not intrinsically quantum mechanical \cite{aa}.  

Previous successful implementations of this intriguing imaging method using visible light \cite{c, d, GI-expt-with-visible-light, GI-expt-with-visible-light-3}, atoms \cite{j} and x-rays \cite{k, l, m} open profound new pathways to imaging. Exploration of these new pathways remains in its infancy.

Ghost imaging retains some level of mystery on account of the aforementioned debate as to whether the process is inherently quantum mechanical (relying, for example, on photon entanglement), or intrinsically classical \cite{aa}.  We concur with the currently-dominant opinion that some ghost imaging scenarios may be understood in purely classical terms, while other forms of ghost imaging rely intrinsically on quantum processes \cite{aa}.  The class of ghost imaging experiment, to which the present paper belongs, may be understood in purely classical terms.   

Ghost imaging has several particularly attractive features: (i) In its computational-imaging variant \cite{c, Shapiro2008, n}, ghost imaging permits images of an object to be formed without any position-sensitive detectors whatsoever. Radiation is only detected using a large one-pixel ``bucket detector'', with the illuminating intensity patterns being known and therefore not needing to be measured \cite{a, c, d, n}. This may permit fuller utilisation of all radiation scattered by a sample of interest, via multiple buckets.  This possibility may be particularly attractive for imaging using X-rays, atoms and neutrons.  (ii) Turbulence robustness \cite{GI-with-turbulence1, GI-with-turbulence2, GI-with-turbulence3, GI-with-turbulence4} is another attractive feature of the method. Since it relies on intensity {\em correlations}, the ghost signal is unaffected by non-correlated random perturbations that perturb the input pair of signals. (iii) Forms of ghost imaging may exist which lead to reduced radiation dose on the sample \cite{l, p}, a possibility which is especially appealing when using ionizing radiation.  One compelling proposal considers x-ray free-electron lasers, utilising  parametrically down-converted photons passing through the object that are entangled with x-ray-photons that never pass through the object \cite{p}.  More broadly, an attractive avenue for further research is to seek ghost-imaging protocols that offer reduced radiation dose to a sample of interest.     

We restrict attention to {\em x-ray} ghost imaging for the remainder of this paper.  The current literature on experimental realisations of x-ray ghost imaging is sparse.  Inspired by earlier proposals \cite{Cheng2004, adams}, two papers were published in 2016, both reporting the experimental realisation of ghost imaging in one transverse dimension using x-ray synchrotron sources \cite{k, l}.  Ghost imaging using a laboratory x-ray source, again in one transverse dimension, was reported in 2017 \cite{m}.  Very recently, and experimental realization of 2D x-ray ghost imaging with a laboratory source was reported \cite{zhang2017}, providing the first experimental evidence of the possibility of dose reduction afforded by the ghost imaging protocol. Even taking into account its evident roots in experimental x-ray studies \cite{q, r, s, t, u, v} on the Hanbury Brown--Twiss (HBT) effect \cite{b1, b2, Wang2009}, the paucity of extant literature on experimental realisations of x-ray ghost imaging is an obvious indicator of rich avenues for future work.       

With the notable exception of the extremely recent manuscript by Zhang {\em at al.} \cite{zhang2017}, all published reconstructions \cite{k, l, m} are one-dimensional.  Note, in this context, that the earlier x-ray HBT studies \cite{q, r, s, t, u} may be viewed, at least in retrospect, as zero-dimensional ghost imaging.  The key point is the dearth of experimental x-ray ghost reconstructions in higher than one dimension.  Another limitation of existing approaches \cite{l} is in the method by which speckle bases are produced not being readily scalable to higher-dimensional and higher-resolution ghost imaging.  Of the currently-published approaches, that of Yu {\em et al.} \cite{k} with a synchrotron x-ray source, Schori and Schwartz \cite{m} and Zhang \textit{et al.} with a tabletop x-ray tube source utilise spatially random masks to generate the ensemble of speckle patterns that form the set of linearly independent basis functions used in the ghost reconstruction.  The use of masks appears particularly amenable to scaling up to both higher-dimension and higher-resolution x-ray ghost imaging.  Another difficulty inherent to current approaches is the very precise timing resolution needed in the approach of Pelliccia {\em et al.} \cite{l}, since this employs the pulsed nature of third generation x-ray synchrotron emission from individual electron bunches to provide an ensemble of speckle patterns as input into the ghost reconstruction.  Note however that use of x-ray speckles produced by pulsed sources is likely to be very promising for SASE free electron lasers, due to their high spatial coherence (and therefore high speckle contrast). 

Here we present a two-dimensional experimental realisation of x-ray ghost imaging, with a higher number of pixels in each transverse dimension than all of the previously cited experimental studies.  We believe this method to be practical in the sense of being scalable to higher dimensional imaging (i.e. tomography) and finer spatial resolutions.  The method is also amenable to highly parallel geometries, in which ghost images of a large number of objects may be acquired simultaneously.  Moreover, our analysis explicitly benefits from the previously mentioned turbulence robustness of the method, by being insensitive to uncorrelated fluctuations between the object and reference arms of our ghost-imaging setup.  

We close this introduction with a brief outline of the remainder of the paper.  Section 2 gives an outline of the x-ray synchrotron-based experimental setup and measurement process for practical two-dimensional ghost imaging.  Section 3 presents our reconstructed two-dimensional images, of both a perforated lead-sheet stencil and the tungsten filament of an incandescent light globe. Two different methods of ghost-image reconstruction are considered, together with a brief study comparing spatial resolution in the two different methods of ghost imaging.  Section 4 comprises a discussion on diverse topics such as a means for determining the point spread function associated with a given ghost-imaging speckle basis, the turbulence robustness of the method, and future possibilities including parallelised x-ray ghost imaging, computational x-ray ghost imaging, and x-ray ghost tomography.      

\section{Experimental realization of x-ray ghost imaging}
\begin{figure}[b]
\begin{center}
\includegraphics[width=8cm]{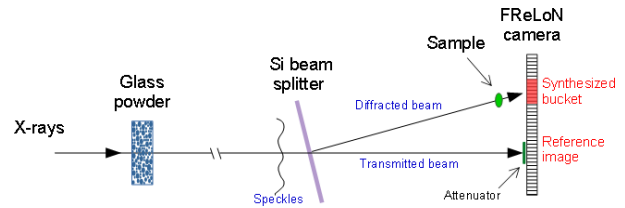}
\end{center}
\caption{Schematic diagram of the experimental setup.}
\label{fig:1}
\end{figure}
\subsection{Experimental setup}
\begin{figure}[t]
\begin{center}
\includegraphics[width=8cm]{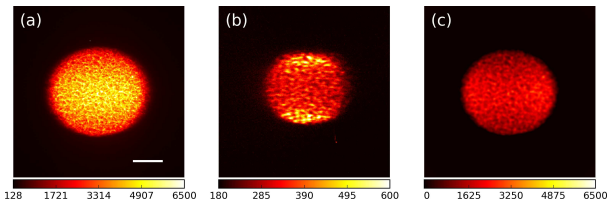}
\end{center}
\caption{(a) Image of the primary beam on the FReLoN camera acquired with 2 s exposure time. The beam was attenuated by stacking a 500 \textmu m thick Cu foil and a 500 \textmu m thick GaAs wafer to protect the camera. Scale bar = 2 mm. (b) Corresponding image of the diffracted beam. No attenuator was placed in the diffracted beam path. (c) Blurred version of the image in (a) to highlight the similarities in the speckle pattern distribution between the direct and the diffracted beam.}
\label{figure:2}
\end{figure}

The experiment was carried out at beamline ID19 of the European Synchrotron ESRF in Grenoble (France). Undulator light with a mean energy of 26.3 keV was focused by a stack of compound refractive lenses to a focal spot of about 5.5 mm diameter at the sample. To produce variable (and controllable) speckles in the beam we inserted a 1 cm thick perspex container filled with glass powder in the beam, 5.8 m upstream of the sample. The powder (Oberfl\"achentechnik Seelmann, Germany) was composed of grains, irregular in shape, with typical size distributed in the range 200--1000 \textmu m. Propagation based phase contrast from the beads generated a speckle pattern on the sample, that could be controlled by raster scanning the glass-beads slab in the transverse plane. A silicon crystal beam splitter, placed 20 cm upstream of the sample, was used to produce two copies of the beam by means of Laue diffraction (transmission geometry) from the (220) planes of the silicon. The primary beam was mostly transmitted by the silicon wafer (see scheme in Fig. \ref{fig:1}), thus creating two non-identical copies of the beam. Both beams were then recorded by the same pixel array detector (scintillator lens-coupled to a FReLoN
camera), placed immediately downstream of the sample. The effective pixel size of the camera was 30 \textmu m. 
The attenuated image of the primary beam, with the glass-beads slab in place, is shown in Fig. \ref{figure:2}(a). An attenuator composed of a 500 \textmu m thick Cu foil and a 500 \textmu m thick GaAs wafer was inserted in the primary beam to avoid saturation and protect the camera during the prolonged exposures. The corresponding image of the diffracted beam (cropped from the same frame of the camera) is shown in Fig.~\ref{figure:2}(b). The image was acquired with 2 s exposure time. Notice the different intensity of the two beams, with the diffracted beam being much weaker compared to the primary beam. The ratio of the average intensities in the center of the direct and diffracted beam was estimated by successive measurements with and without attenuators to be $1.4 \times 10^{-4}$. 

Two main features of the diffracted beam are worth remarking. (i) Due to unavoidable vibration of the silicon crystal, the image of the diffracted beam appears blurred when compared to the primary beam. To facilitate comparison between the two, we show the blurred version of the primary beam image in Fig.~\ref{figure:2}(c). The blurring was performed using a Gaussian kernel with $\sigma=60$  \textmu m (2 pixels). The size of the speckles in the blurred image is comparable with the corresponding size of the speckles of the diffracted beam, so that the similarity between the two is more evident. (ii) Likely due to some strain generated in the crystal mount, the diffracted beam appears to be slightly compressed, with  more intensity being diffracted around the top and bottom edge of the beam. This led, as we will see in Sec. 3.2, to a slight demagnification of the ghost images when compared to the original images.     

\subsection{Ghost imaging measurement procedure}
The sample was inserted in the diffracted beam.  We decided on this configuration to ensure that the sample was being illuminated by the weaker beam. Given our estimated intensity ratio, and assuming the measured counts to be proportional to the number of photons (we are in a nearly monochromatic case), we expect that the sample receives a fraction of 0.014\% of the photons that are in the reference beam.  The variable speckle illumination was obtained by raster scanning the glass-beads slab in the transverse plane. The range of the raster scan was 150 mm $\times$ 90 mm (H $\times$ V), with a step size of 750 \textmu m $\times$ 500 \textmu m  (H $\times$ V). These parameters were chosen to ensure the step size to be about 2 times larger than the typical speckle size (compatible with the total size of the glass beads sample), so that images taken at neighbouring positions along the scan were nearly independent. 

A speckle image was acquired at each position of the glass-beads slab, with 2 s exposure time. The bucket signal was then synthesized by summing the number of counts in an area of 200 $\times$ 180 pixels comprising the diffracted beam. The reference image at each position was generated by cropping an area of 260 $\times$ 230 pixels centered around the primary beam from the raw (attenuated) camera image. We acquired a total of 5000 frames for each ghost imaging reconstruction.  

Due to the periodic electron injections into the ESRF storage ring, a low frequency time structure (with period of 1 h) was present in the intensity signal. To prevent this feature affecting the ghost imaging reconstruction, the data (both bucket signal and reference images) were Fourier filtered to remove such low frequency components. The filter was a simple low-pass designed to remove all frequency below 5\% of the Nyquist (temporal) frequency, and therefore discard all slow variations of the beam intensity due to the injections.

\section{Ghost imaging reconstruction results}

\subsection{Ghost imaging reconstruction strategy}
Conventional ghost imaging reconstruction can be obtained with the superposition formula \cite{c,d}:
\begin{equation}
v_{i} = \frac{1}{m} \sum_{j=1}^{m} \left(b_j - \bar{b} \right)A_{ij}. 
\label{eq:randomGI}
\end{equation}

In Eq. (\ref{eq:randomGI}), $v_{i}$ is the \textit{i}-th pixel of the (rasterized) ghost image \textbf{v},  written as the superposition of the corresponding pixels of the measured speckle images $ A_{ij} $ (\textit{i}-th pixel of the \textit{j}-th measurement). Each term of the superposition is weighted by the corresponding bucket signal $b_j $ subtracted by its mean $\bar{b}$. The total number of images used in the process is \textit{m}. In  our experiment $m=5000$. The ghost image $\mathbf{v}$ consists of $n = n_1 \times n_2$ pixels. Denoting by $\mathbf{A}$ the $m \times n$ matrix of reference images, the bucket signal is $\mathbf{b} = \mathbf{A}\mathbf{v}$.

Equation (\ref{eq:randomGI}) can be written in the compact form 
\begin{equation}
\mathbf{v} = \langle  \left( \mathbf{b} - \langle \mathbf{b} \rangle \right) \, \mathbf{A} \rangle, 
\label{eq:randomGI2}
\end{equation}
where the symbol $\langle \rangle$ denotes ensemble average. 

The use of a random measurement matrix $\mathbf{A}$ however, is not optimal: to attain a good signal-to-noise ratio (SNR) a large number of measurements is generally required ($m\gg n$), which makes the basic protocol unsuitable for applications demanding low dose. As noted in \cite{d}, the reason for that becomes apparent by interpreting Eq. (\ref{eq:randomGI}) as a conventional expansion using linearly independent basis functions. The rows of the matrix $\mathbf{A}$ represent the basis vectors, and the process of ghost imaging is in fact a projection operation of the image on such a basis. A better basis is made of orthonormal rather than merely linearly independent vectors, and therefore the quality of the ghost image recovery depends on how well the rows of the random measurement matrix approximate an orthonormal basis---see Appendix for more detail on this point.  Further to this, it is also worth noting that genuinely random illumination can be hard to produce in practice. In our case, we took care to scan the glass-beads slab by a transverse step size that was much larger than the transverse speckle size, however residual correlations are still present. 

To overcome such limitations causing a non optimal choice of the measurement matrix, several approaches are currently used. Compressive ghost imaging speeds image recovery using ideas and techniques of compressive sensing \cite{d}. Image recovery can be improved by compressive sensing by identifying an orthonormal basis in which the image to be recovered is \textit{sparse} \cite{d,candes06,candes08}. 

In an alternative approach, commonly used in single pixel cameras, image recovery can be much improved if one starts from an orthonormal measurement matrix (or sensing matrix) in the first place. A common choice is the Hadamard matrix  $\mathbf{H}$  implementing Hadamard--Walsh functions via a spatial light modulator (see for instance \cite{clemente13}). In this case, redefining the bucket signal as $\mathbf{b}_H = \mathbf{H}\mathbf{v}$, in the absence of noise, the expansion in Eq. (\ref{eq:randomGI}) is exact when $m=n$. While this is a convenient choice in the visible part of the spectrum (where efficient spatial light modulators exist), this option is not easy to implement for x-ray imaging. 

Our approach was to assume that the speckle illuminating function is approximately orthogonal, and perform an effective orthogonalization using the QR decomposition of the matrix $\mathbf{A}$:
\begin{equation}
\mathbf{A} = \mathbf{Q}\mathbf{R}.
\label{qr}
\end{equation}

\noindent Here, $\mathbf{Q}$ is the orthogonal matrix we seek, and $\mathbf{R}$ is an upper triangular matrix. 

The only non-trivial step is to rearrange the bucket vector $\mathbf{b}$ accordingly. Specifically we seek a vector $\mathbf{\tilde{b}} = \mathbf{Qv}$, that is the bucket signal that would be measured if the measurement matrix were the orthogonal matrix $\mathbf{Q}$. Since $\mathbf{b} = \mathbf{A}\mathbf{v}$, we can write

\begin{equation}
\mathbf{b} = \mathbf{QRv} = \mathbf{QRQ}^{-1}\mathbf{Q}\mathbf{v} =\mathbf{AQ}^{-1}\mathbf{\tilde{b}} .
\label{bgs1}
\end{equation}
Therefore, by inverting the previous expression:
\begin{equation}
\mathbf{\tilde{b}} = \mathbf{QA}^{-1}\mathbf{b}.
\label{bgs2}
\end{equation}

Equations (\ref{qr}) and (\ref{bgs2}) constitute the algorithm we use on our data to produce an approximately orthonormal decomposition. The ghost imaging reconstruction can then be obtained using Eq. (\ref{eq:randomGI2}) with $\mathbf{b} \rightarrow \mathbf{\tilde{b}}$ and $\mathbf{A} \rightarrow \mathbf{Q}$ (note that $\langle  \mathbf{\tilde{b}} \rangle =0$): 

\begin{equation}
\mathbf{v} = \langle  \mathbf{\tilde{b}} \, \mathbf{Q} \rangle.
\label{gigs}
\end{equation}
\begin{figure}[b]
\centering\includegraphics[width=8cm]{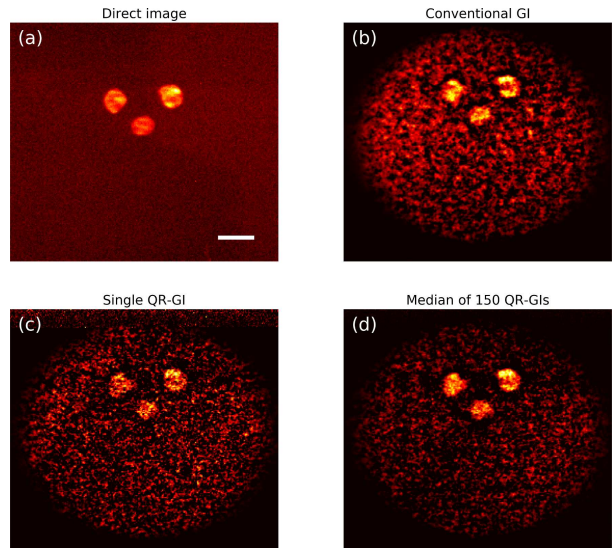}
\caption{Measurement of the stencil sample. (a) Direct image of the sample when illuminated by one realization of the speckle pattern. (b) Conventional ghost imaging reconstruction using $m=5000$ measurements. (c) Ghost imaging reconstruction after QR decomposition of the measurement matrix, obtained using the same measurement as the previous case. (d) Median image of 150 ghost images obtained by QR decomposition of the randomly permuted measurement matrix. The scale bar corresponds to 1 mm.}
\label{fig:3}
\end{figure}

Using Eq. (\ref{gigs}) for the reconstruction has two main advantages over the conventional superposition formula in Eq. (\ref{eq:randomGI2}). First, it guarantees optimal use of the information, as the new measurement matrix is now composed of orthogonal rows. Second, since the QR decomposition scrambles the basis, the typical speckle size of the orthogonal measurement matrix $\mathbf{Q}$ become effectively smaller than the typical size of the physical speckles used in the measurement. This means that the resolution of the reconstructed ghost image is no longer limited by the real speckle size, at the price of an increased noise in the reconstruction. 

Notably, one could also overcome the noise problem, by noting that for underconstrained problems ($m<n$) one could perform a QR decomposition multiple times by permuting the rows of $\mathbf{A}$ (and, correspondingly, of $\mathbf{b}$) each time. In practice one obtains a different reconstruction each time, but the difference in the reconstructions is mostly in the noise background. By averaging multiple reconstructions (using always the same data, hence not increasing the radiation dose), one could reduce the noise and increase the resolution of the reconstruction at the same time. 

\subsection{Experimental results}

\begin{figure}[b]
\begin{center}
\includegraphics[width=8cm]{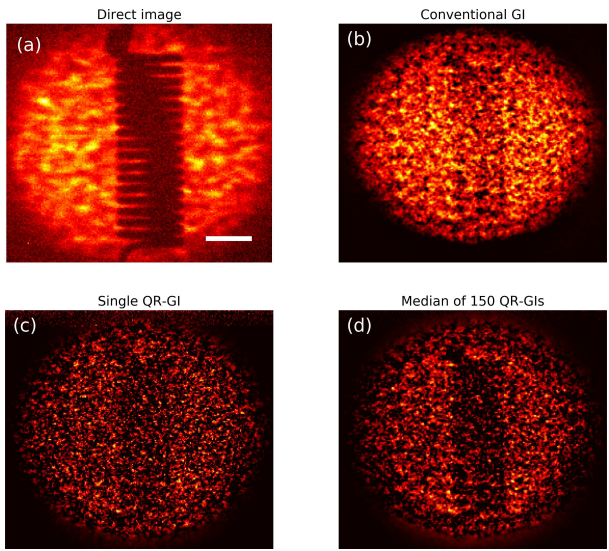}
\end{center}
\caption{Measurement of the tungsten coil. (a) Direct image of the sample when illuminated by a realization of the speckle pattern. (b) Conventional ghost imaging reconstruction using $m=5000$ measurements. (c) Ghost imaging reconstruction after QR decomposition of the measurement matrix, obtained using the same measurement as the previous case. (d) Median image of 150 ghost images obtained by QR decomposition of the randomly permuted measurement matrix. The scale bar corresponds to 1 mm.}
\label{fig:4}
\end{figure}

The first sample we imaged was a stencil obtained by drilling three holes in a lead sheet. The direct image of the sample in the diffracted beam (before synthesizing the bucket signal) is shown in Fig.~\ref{fig:3}(a). The ghost image, obtained using the conventional reconstruction formula in Eq. (\ref{eq:randomGI2}) with $m=5000$, is shown in Fig. \ref{fig:3}(b). Note that the ghost image size is equal to the reference image size of $260 \times 230 = 59800$ pixels. Therefore we acquired a little more than 8\% of the Nyquist sampling. The ghost imaging reconstruction clearly reproduces the sample features albeit at reduced resolution, as dictated by the speckle size. Significant background noise is also present as a consequence of the limited number of measurements and the camera noise. 

As discussed in the previous section, resolution can be improved via QR decomposition of the measurement matrix. The result of this operation (see Eqs. (\ref{bgs1}) -- (\ref{gigs})) is shown in \ref{fig:3}(c). The resolution of this image is much improved, to the detriment of the image noise which is increased. Incidentally, this noise--resolution trade-off is consistent with the noise--resolution uncertainty principle, recently introduced by Gureyev \textit{et al.} \cite{gureyev2016}. Conversely, by repeating the QR decomposition 150 times (each time performing a random permutation of the rows of the measurement matrix $\mathbf{A}$ and, correspondingly, of the bucket values $\mathbf{b}$) and taking the median of those images, the map in Fig. \ref{fig:3}(d) can be synthesized. This last image is comparable to the conventional reconstruction in terms of noise, and displays higher resolution. 

The same set of images for the second sample is shown in Fig. \ref{fig:4}. In this case the sample was a tungsten coil. The map obtained by the median of 150 ghost images obtained after QR decomposition shows a marked improvement over the conventional ghost image, which reflects the advantage of the QR decomposition method in optimising the use of the available information.

When compared to the stencil reconstruction, the ghost image of the coil looks noisier. Both images have been reconstructed using the same number of measurements. The difference is to be found in the sample extent compared to the beam size. The stencil sample is effectively composed of three holes only, whose size is relatively small compared to the beam. That means that, variations in the speckles position amount to relatively large excursions of the buckets signal. Conversely, the bucket signal after the coil sample will vary comparably much less, as it receive contribution from most of the beam size. Therefore we expect that the ghost imaging procedure is much more sensitive when reconstructing the stencil, as opposed to the tungsten coil. 

Finally, as anticipated in Sec. 2.1, due to the beam compression effected by the silicon beamsplitter, the ghost images appeared demagnified when compared to the original images. This is especially evident observing the coil images in Fig. \ref{fig:4}, while however being present to the same extent in the ghost image reconstructions of both samples.

\section{Discussion}
\subsection{Completeness relation and Point Spread Function of the ghost imaging system}

The standard ghost imaging formula in Eq. (\ref{eq:randomGI}) considers the ensemble of linearly independent random speckle images as a basis from which to synthesize the reconstruction.  Indeed, as explored in more detail in the Appendix, the standard ghost imaging formula may be viewed as a superposition of approximately orthogonal functions.  A direct consequence is that the rows of the measurement matrix \textbf{A} should obey an ``approximate completeness relation'', which can be written as:

\begin{equation}
\frac{1}{m}\sum_{j=1}^m \left( A_{ij}-\bar{A} \right) \left( A_{kj} -\bar{A} \right) \approx \delta_{jk} 
\label{eq:complet1}
\end{equation}
where $\delta_{jk}$ is the Kronecker delta. Note that, subtracting the average $\bar{A}$ from each coefficient is required to have zero-mean terms, as each of the $A_{ij}$ is non-negative on account of it being a measured intensity value. The previous equation would be exact only in the ideal case in which the measurement matrix (after subtracting its average) forms a complete orthonormal set.

To make the previous argument more apparent, let us explicitly rewrite the rows of the measurement matrix as the measured speckle images $I_j(x,y)$, where $x,y$ are the coordinates on the detector plane and index $j$ runs over the number of measurements. With this new, more transparent, notation the completeness relation in Eq. \ref{eq:complet1} can be rewritten as: 
\begin{equation}
\begin{split}
\frac{1}{m} \sum_{j=1}^m \left[ I_j(x,y) - \bar{I} \right] \left[ I_j(x',y') - \bar{I} \right] \\
\approx \delta(x-x',y-y'),
\end{split} 
\label{eq:complet2}
\end{equation}
where $\delta(x,y)$ is the Dirac delta. As before, Eq. (\ref{eq:complet2}) is exact only when the speckle images subtracted by their average form a complete orthonormal set. In all practical cases the previous equation can be used instead to define an effective Point Spread Function (PSF), $ \delta(x-x',y-y') \rightarrow \textrm{PSF}(x-x',y-y')$ centered around the point $x', y'$ (see Eq. (\ref{eq:App8}) in the Appendix for a more detailed analysis on this point).

\begin{figure}[t]
\begin{center}
\includegraphics[width=8cm]{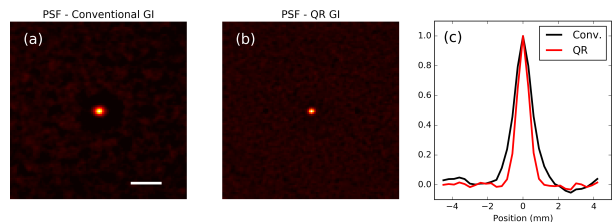}
\end{center}
\caption{(a) PSF of the ghost imaging system, calculated using Eq. (\ref{eq:complet2}), for the conventional ghost imaging situation. The scale bar represents 500 \textmu m. (b) Corresponding PSF calculated after QR decomposition. The PSF appears noticeably narrower, reflecting the resolution improvement afforded by the QR decomposition. (c) Line profile taken across the central horizontal line in the maps in (a) -- black solid line -- and (b) -- red solid line. When fitted with a Gaussian function, the two peaks have a FWHM of 125 \textmu m  and 80 \textmu m respectively.}
\label{fig:5}
\end{figure}

Such a PSF defines in fact the spatial resolution of the ghost imaging system. In the light of this idea, we calculated Eq. (\ref{eq:complet2}) for the conventional ghost imaging situation (where the $I_j(x,y)$ are the original speckle images) as well as for the modified speckle images after the QR decomposition. The results are shown in Fig. \ref{fig:5}. When using the original mask the FWHM of the PSF (fitted with a Gaussian function) turns out to be about 125 \textmu m, which reduces to about 80 \textmu m after the QR decomposition. 

The form of ghost imaging presented here -- equipped with the definition of the PSF as discussed above -- may be therefore viewed as a form of scanning probe imaging \cite{STEM} using a completely delocalized probe and a large integrating bright-field detector.  While the resolution of scanning probe imaging is usually dictated by the size of a localised scanning probe, for our delocalized scanning probe the resolution is limited by the smallest characteristic length scale present in the intensity fluctuations of the ensemble of illuminating 
speckle fields (cf. Eq.(\ref{eq:App8}) in the Appendix), which is in turn the width of the PSF calculated using Eq.
(\ref{eq:complet2}).  
\subsection{Other remarks on future practical aspects of x-ray ghost imaging}
We have previously mentioned the turbulence robustness of ghost imaging \cite{GI-with-turbulence1,GI-with-turbulence2,GI-with-turbulence3,GI-with-turbulence4}.  This turbulence robustness arises from the invariance of the ensemble average in Eq. (\ref{eq:randomGI2}), with respect to the addition of statistically uncorrelated fluctuations (``turbulence'') in the object and reference arms of the ghost-imaging setup.  Stated more precisely, the ensemble average in Eq. (\ref{eq:randomGI2}) is unchanged under either or both of the replacements ${\bf b}\rightarrow {\bf b}+{\delta \bf b}$ and ${\bf A}\rightarrow {\bf A}+{\delta \bf A}$, where $\delta \bf b$ is a zero-average random fluctuation added to the bucket signal and $\delta \bf A$ is a random fluctuation added to the speckle images, provided that $\delta \bf b$ and $\delta \bf A$ are not correlated.  This robustness enhances the practicality of the method.

\begin{figure}[t]
\begin{center}
\includegraphics[width=8cm]{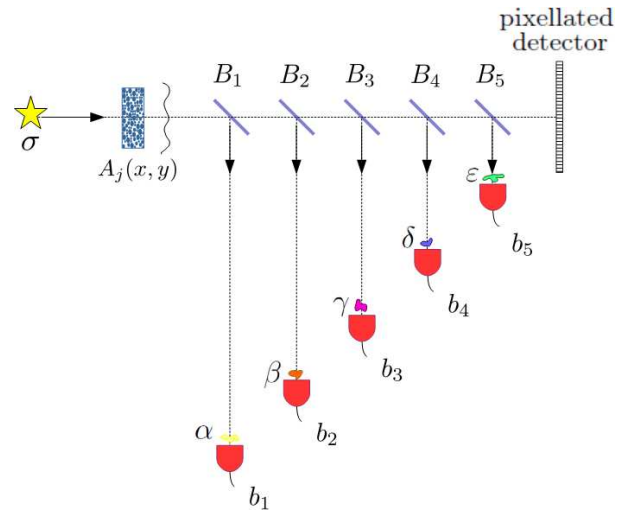}
\end{center}
\caption{Schematic setup for parallelised x-ray ghost imaging.}
\label{fig:6}
\end{figure}

Another practical aspect, of x-ray ghost imaging in the experimental setup used here, is its ability to be parallelised.  Inspired by the parallel form of computational ghost imaging proposed by Yuan {\em et al.} \cite{Yuan2016}, consider the setup shown in Fig.~\ref{fig:6}.  Here, an x-ray source $\sigma$ illuminates an ensemble of $m$ random speckle-producing masks $\{A_j(x,y)\}$, where $j=1,\cdots,m$ labels each realisation of the mask and $(x,y)$ are coordinates in the plane perpendicular to the optic axis.  A series of beam-splitters $B_1, B_2, \cdots$ then illuminate a series of objects $\alpha, \beta, \cdots$, giving associated non-spatially-resolved signals in the bucket detectors $b_1, b_2, \cdots$  Each bucket signal in each detector may be correlated with the same ensemble of speckle images registered by the pixellated array detector for each realisation of the mask, to yield independent parallelised ghost imaging. The objects in in Fig.~\ref{fig:6} are staggered so as to keep constant the source-to-object distance, thereby ensuring that Fresnel diffraction and other free-space-propagation effects are accounted for, with the registered speckle pattern measured over the pixellated array detector being equal (up to a multiplicative constant) to the speckle patterns illuminating each object.  Note that each beamsplitter only needs to remove a negligible fraction of the total energy from the beam which ultimately illuminates the pixellated array detector; the resulting attenuation of the speckle-basis images registered by the array detector can be trivially taken into account in the parallel ghost reconstructions.  Note also that, while Fig.~\ref{fig:6} indicates one object per beam-splitter, one could also have multiple objects per beam-splitter, using multiple Bragg or Laue reflections from a crystal beam-splitter, or multiple Laue reflections from a polycrystal beam-splitter.

Irrespective of whether or not the ghost-imaging geometry is parallelised, one can also consider x-ray ghost imaging scenarios which reduce or even eliminate the reliance on a position sensitive detector.  With this end in mind, suppose that the ensemble of spatially random screens in Figs~1 
or \ref{fig:6} may be generated in a reproducible fashion using a radiation-hard mask.  Such masks might be generated by transversely displacing a highly structured object (such as the layer of glass beads used in the present study), and also by rotation of a suitable highly structured object (such as a thick metallic foam). In this scenario, {\em one need measure the ensemble of reference speckle fields only once}---the pixellated array detector can then be switched off, or even removed, and ghost images subsequently recorded using only the bucket detectors, by running the mask through the previously mentioned reproducible set of positions/orientations.  Note, moreover, that the acquisition time in this geometry can be driven down significantly, since there are no longer any position-sensitive x-ray detectors, together with their associated readout times. Note also that the same set of speckle images could be used with each orientation of an object, in a tomographic x-ray ghost imaging context.   

Interestingly, in the spirit of computational (ghost) imaging \cite{c, Shapiro2008, n}, one may even dispense altogether with the pixellated array detector.  To do this, one would have a highly structured mask whose three-dimensional micro-structure is so well characterised, and the illuminating beam so stable and well characterised, that one can use a numerical implementation of the x-ray scattering and diffraction to calculate the ensemble of reference speckle fields that one would have measured had an array detector been used; these images therefore do not need to be measured.  One would then have a form of {\em computational x-ray ghost imaging} using only bucket detectors.  In this context, we point out that the function played by spatial light modulators in visible-light computational imaging is replaced with the known micro-structured mask, in our proposed form of x-ray computational imaging.  In the near future, this highly structured mask might even be amenable to fabrication using three-dimensional printers.  

We close the discussion with a final remark: spatially random masks are not necessarily optimal for x-ray ghost imaging.  While random masks have the virtue of ease of synthesis,  structures such as the uniformly redundant array \cite{URA} may be more efficient.

\section{Conclusion}
We have presented a practical realization of x-ray ghost imaging using synchrotron x-rays from an undulator. This experimental demonstration shows a practical avenue for producing x-ray ghost images. We reported the measurements of two samples, a stencil in a lead mask and a tungsten coil. For both samples we reported two different reconstruction strategies. The first is based on the conventional ghost imaging formula, in which the ghost image is approximated by the weighted average of the speckle illuminating images. The weights of the superposition are the bucket signals subtracted by their average. The second approach is based on prior QR decomposition of the measured speckle reference images. In this way the ensemble of speckle images can be made to be a better approximation to an orthogonal basis, thereby improving the resolution of the reconstruction.  Next, we analyzed in more detail the resolution of our ghost imaging system, defining an effective Point Spread Function (PSF) which was shown to be improved upon QR decomposition of the illuminating functions. Finally, we discussed practical aspects for future applications of x-ray ghost imaging, including its robustness against turbulence and improved measurement strategies using parallelised ghost imaging and computational x-ray ghost imaging.

\section*{Acknowledgments}
We thank the directors of the ESRF for funding DP and DMP to visit in early 2017.  S. B\'{e}rujon and E. Brun loaned us the speckle masks used in our experiment. T.E. Gureyev suggested we investigate the completeness relation in the context of the present work.  We acknowledge useful discussions with T.E. Gureyev, I.D. Svalbe, G.R. Myers, A. Kingston and D. Ceddia.

\section*{Appendix: Estimating ghost-imaging resolution from a given speckle basis}

Consider an ensemble of $m \gg 1$ spatially random two-dimensional intensity speckle patterns $\{I_j(x,y)\}$ defined over a domain $\Omega$ with area $A(\Omega)$ in the $(x,y)$ plane.  Each ensemble member is labelled by the integer $j$, with each $I_j(x,y)$ being non-negative on account of being an intensity map.  

Assume ``many speckles'' in each random speckle field.  More precisely, consider each to be a distinct realisation of a stochastic process with every member of the ensemble of speckle fields having the same characteristic transverse length scale $\sigma$, in both $x$ and $y$.  Stated differently, we assume each realisation of the speckle field to be statistically identical.  Since each speckle field has $M$ speckles with $M \approx A(\Omega)/\sigma^2 \gg 1$, this implies that (i) the spatially-averaged intensity of each realisation of the ensemble is approximately the same, (ii) the spatially averaged squared intensity, of each member, is also approximately the same, and (iii) the ensemble-averaged intensity at any point in $\Omega$ is independent of position, and approximately equal to the spatially averaged intensity in any particular realisation. 

The standard ghost-imaging formula considers the ensemble $\{I_j(x,y)\}$ as a speckle basis, from which the ghost image of the object transmission function $v(x,y)$ may be synthesized \cite{c,d}:

\begin{equation}
v(x,y) \circledast \textrm{PSF}(x,y) = \frac{1}{m} \sum_{j=1}^{m}(b_j-\overline{b})I_j(x,y).
\label{eq:App1}
\end{equation}

\noindent Here, $\circledast$ denotes convolution over $x$ and $y$, $\textrm{PSF}(x,y)$ is a point spread function associated with the finite spatial resolution with which $v(x,y)$ is estimated, the bucket signal is

\begin{equation}
b_j=\iint_{\Omega}v(x,y)I_j(x,y) dx dy
\label{eq:App2}
\end{equation}

\noindent and the average bucket signal is

\begin{equation}
\overline{b}=\frac{1}{m}\sum_{j=1}^m b_j.
\label{eq:App3}
\end{equation}

Using the above definitions and assumptions, one can readily show that

\begin{equation}
\overline{b}=\overline{I}~\overline{v} A(\Omega),
\label{eq:App4}
\end{equation}

\noindent where 

\begin{equation}
\overline{I}=\frac{1}{m}\sum_{j=1}^mI_j(x,y)\left(=\frac{1}{A(\Omega)}\iint_{\Omega} I_j(x,y) dx dy \textrm{ for any }j\right),
\label{eq:App5}
\end{equation}

\noindent and 

\begin{equation}
\overline{v}=\frac{1}{A(\Omega)}\iint_{\Omega} v(x,y) dx dy.
\label{eq:App6}
\end{equation}

Using Eqs. (\ref{eq:App2}), (\ref{eq:App4}) and (\ref{eq:App5}), Eq. (\ref{eq:App1}) can be manipulated into the form:

\begin{equation}
\begin{split}
v(x,y)\circledast \textrm{PSF}(x,y) \\
= \iint_{\Omega} dx' dy' v(x', y') \frac{1}{m}\sum_{j=1}^m[I_j(x',y')-\overline{I}][I_j(x,y)-\overline{I}].
\end{split}
\label{eq:App7}
\end{equation}

Since $v(x,y)\circledast \textrm{PSF}(x,y) = \iint_{\Omega}v(x',y')~\textrm{PSF}(x-x',y-y') dx' dy'$, the point-spread function associated with the ghost-imaging reconstruction is the ensemble-averaged intensity--intensity correlation between the locations $(x,y)$ and $(x',y')$, namely the smoothed completeness relation:

\begin{equation}
\textrm{PSF}(x-x',y-y') = \frac{1}{m}\sum_{j=1}^m[I_j(x',y')-\overline{I}][I_j(x,y)-\overline{I}].
\label{eq:App8}
\end{equation}

For any fixed position $(x,y)=(x',y')$, and a specified measured ensemble of intensity-speckle fields, Eq. (\ref{eq:App8}) can be used to estimate the local resolution associated with superpositions (such as Eq. (\ref{eq:App1})) that utilise this basis in a ghost imaging context---see Fig.~\ref{fig:5}.  The above expression, the right side of which is the same correlation function (intensity covariance) measured in classic Hanbury Brown--Twiss scenarios \cite{b1,b2}, may also be used to generate a ``pixel basis'' consisting of a lattice of PSFs whose centroids are separated by the full-width-at-half-maximum of each PSF.  This set of PSFs is another approximately orthogonal basis, which is complementary to the approximately-orthogonal speckle basis from which it is derived, insofar as the former is localised whereas the latter is not.  Indeed, one may loosely speak of Eq. (\ref{eq:App8}) as showing how linear combinations of the spatially delocalized speckle probes may be formed to yield very localised probes.    

Interestingly, a variant of the above chain of reasoning may be used to {\em{derive}} the standard ghost-imaging formula from first principles, in a physically transparent manner. One can start with an ensemble of intensity speckle fields that obey the previously articulated assumptions (see second paragraph of this Appendix).  The intensity covariance on the right side of Eq. (\ref{eq:App8}) will typically be a narrowly-peaked normalisable function, making it natural to adopt it as a point-spread function, by definition. The convolution of this point-spread-function with a well-behaved but otherwise arbitrary function $v(x,y)$, then leads directly to the standard ghost-imaging formula in Eq. (\ref{eq:App1}), by reversing the logical development of this Appendix.

We close by noting that the standard ghost imaging formula may be viewed in approximate terms as a form of orthogonal function expansion.  Indeed, for our ensemble of speckle images, each of which are statistically identical and each of which contain many speckles, one has 

\begin{equation}
\frac{1}{A(\Omega)}\iint_{\Omega} \tilde{I}_j(x,y)\tilde{I}_{j'}(x,y) dx dy \approx \delta_{jj'} \textrm{Var}(I),
\label{eq:Appendix2}
\end{equation}

\noindent where $\tilde{I}_j(x,y)$ are background-subtracted speckle fields with

\begin{equation}
\tilde{I}_j(x,y)\equiv I_j(x,y)-\overline{I},
\label{eq:Appendix1}
\end{equation}

\noindent and $\textrm{Var}(I)$ is the variance of each $I_j(x,y)$, which is assumed to be independent of $j$ on account of the previously articulated assumptions.  By construction, the set $\{\tilde{I}_j(x,y)\}$ of speckle patterns has each member averaging to zero over $\Omega$, with distinct members being orthogonal to one another, and all members being approximately orthogonal to a constant offset.  Thus the standard ghost imaging formula, as a weighted superposition of the near-orthogonal set of functions $\{\{\tilde{I}_j(x,y)\},1\}$, may be viewed in approximate terms as a form of orthogonal function expansion.


\end{document}